\documentclass[aps,prl,twocolumn,preprintnumbers,groupedaddress]{revtex4}

\usepackage{graphicx} 

\begin{document}


\title{Fourier imaging study of efficient near-field optical coupling in solid immersion fluorescence microscopy}

\author{Masahiro Yoshita, Kazuko Koyama, Motoyoshi Baba, and Hidefumi Akiyama}

\affiliation{Institute for Solid State Physics, University of Tokyo,\\
5-1-5 Kashiwanoha, Kashiwa, Chiba 277-8581, Japan\\}

\date{\today}

\begin{abstract} 
We measured images and Fourier images of fluorescence for 0.11- and 0.22-$\mu$m-diameter dye-doped polystyrene micro-sphere beads on a solid immersion lens, and experimentally verified strongly-angle-dependent fluorescence intensities due to efficient near-field optical coupling in solid immersion fluorescence microscopy. The results are interpreted in comparison with calculated emission patterns of an emission dipole placed near a solid surface, which establish a basic model for high-collection efficiency in solid-immersion fluorescence microscopy. 
\end{abstract}

\maketitle

A solid immersion lens (SIL) is an aberration-free truncated-sphere-shaped solid lens with high refractive index $n$ used in proximity of a sample for high-resolution near-field optical microscopy \cite{Mansfield90a}. Since 1990, various applications of SILs to high-resolution optical read-out have been reported, such as scanning optical microscopy \cite{Mansfield90a,Ghislain98a,Ghislain99a}, optical storage \cite{Mansfield93a,Terris94a,Terris96a,Ichimura97,Stotz97a,Chekanov99a,Kino99a}, and spectroscopic imaging of semiconductor nano-structures \cite{Sasaki97,Yoshita98a,Yoshita98b,Yoshita00a,Poweleit98a,Wu99,Vollmer99a} and dye molecules \cite{Koyama99,Wu99b,Wu00}. 

Recently, it is found in SIL fluorescence microscopy experiments on semiconductors \cite{Sasaki97,Yoshita98a} and dye molecules \cite{Koyama99} that the SIL fluorescence microscopy has a significant advantage of high collection efficiency simultaneously with high resolution. 

To interpret the high collection efficiency in SIL fluorescence microscopy \cite{Koyama99}, we made numerical calculation of radiation patterns for an emission dipole located in the near-field regime of an air-SIL interface assuming a theoretical model by Hellen and Axelrod \cite{Hellen87}. Calculated radiation patterns show that fluorescence is predominately emitted in the direction of the SIL. 
We estimated collection efficiencies on the basis of the emission patterns under various experimental conditions, and gave an explanation of the observed improvement of collection efficiency in the SIL microscopy. 

However, quantitative experiments to prove the high efficiency in SIL fluorescence microscopy are still few, and the above model to estimate the efficiency seems to require more experimental justifications. For this purpose, efficiency measurements for various combinations of SILs and samples with good accuracy are necessary. Furthermore, direct observation of the radiation patterns of fluorescence in SIL fluorescence microscopy is essentially important. 

We verify, in this paper, the radiation patterns of fluorescence by measuring angle-dependent fluorescence intensities via Fourier imaging as well as angle-integrated fluorescence intensities from both sides of the air-SIL interface. In the angle-integrated-intensity measurement we made statistical analysis of data to accurately evaluate the improvement in collection efficiency. 
These experiments confirm that the radiation patterns predicted by the above dipole model are indeed realized, and that the model well describes the efficient near-field optical coupling in SIL fluorescence microscopy. The results justify the estimation of absolute values of collection efficiency in SIL fluorescence microscopy, and a value of 62\% is obtained in the present experiment.

The samples used in the experiment were dye-doped polystyrene micro sphere beads (Molecular Probes, F-8887) of 0.11 and 0.22 $\mu$m diameters with fluorescence peak at 600 nm. We attached the beads directly to flat surfaces of SILs by putting small drops of water solution of the beads with about $10^{-7}$ concentration in volume and letting water dried out. Though the bead samples are not ideal for theoretical modeling, they are experimentally easy to handle and stable in fluorescence intensity and in spatial position. The SILs used here were 2-mm-diameter hemisphere SILs with refractive index $n$=1.845 (LaSF9-glass) and 1.687 (SF8-glass) prepared by polishing sphere lenses into hemisphere shape (by Ogura Jewel Industry Co., Ltd.). 

We used a standard Nikon optical microscope (Nikon, Optiphot-100s), equipped with an infinite conjugate objective lens with NA=0.8, K\"ohler epi-illumination optics for 546 nm line of 100W mercury lamp, a tube lens, and a cooled charge-coupled device (CCD) camera  (Princeton Instruments, TEA/CCD512TKM/1) for image detection.

\begin{figure}
\includegraphics[width=.4\textwidth]{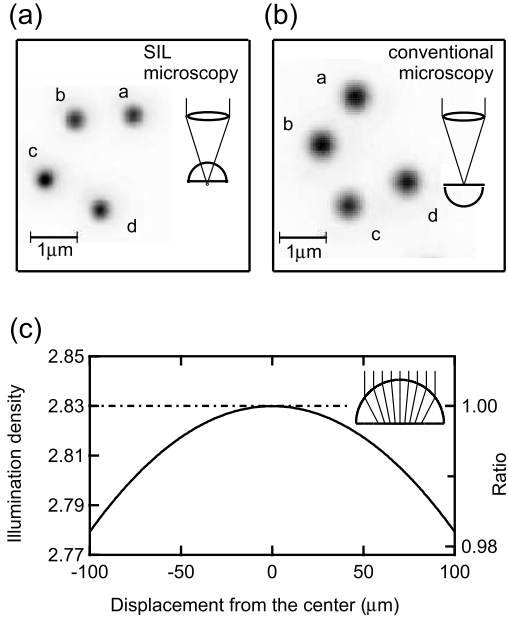}
\caption{
(a) Fluorescence image of 0.11-$\mu$m-diameter beads obtained in the SIL microscopy configuration ($n$=1.845) shown in the inset, and (b) the image obtained in the conventional microscopy configuration shown in the inset. Both images are shown in the common length scale.
(c) Ray traces of K\"ohler illumination through the SIL (inset) and illumination density at a bottom flat surface of a SIL ($n$=1.845).
}
\label{1}
\end{figure}

To perform fluorescence microscopy with a SIL, we set the SIL with beads on the microscope stage with its flat surface down, as shown in the inset of Fig. 1 (a), 
and viewed the beads through the SIL. 
As a reference, we performed conventional fluorescence microscopy, by setting the 
same SIL with beads with its flat surface up, as shown in the inset of Fig. 1 (b), 
using the SIL merely as a glass substrate. 
In the rest of this paper, we denote the measurement in the configuration of the inset of Fig. 1 (a) shortly as the SIL microscopy or the measurement via SIL, while that of Fig. 1 (b) as the conventional microscopy or the measurement via air.

Figures 1 (a) and (b) show the fluorescence images of 0.11-$\mu$m-diameter beads in the same region obtained via SIL ($n$=1.845) and via air, respectively, where we found fluorescence spots of four beads (a-d) in the same arrangement. Spot sizes in Fig. 1 (a) are small, 0.34 $\mu$m on average, compared with those in (b), 0.46 $\mu$m, which demonstrates improved spatial resolution in the SIL microscopy. Figure 1 (a) and (b) are displayed in their respective intensity scales, but actually the spots in (a) were much brighter than those in (b). 

The uniformity and strength of illumination should be commented here. Since in the SIL microscopy we simply added the SIL to the setup of the K\"ohler illumination microscope, illumination rays are refracted at the SIL's surfaces. Figure 1 (c) shows the illumination density in the SIL microscopy ($n$=1.845) in comparison with the original K\"ohler illumination density, which is calculated on the basis of ray tracing schematically shown in the inset. The illumination density is very uniform, within 2\% gradual variation, in the region of 100-$\mu$m-radius around the center, which completely covers the effective field of view for the 2-mm-diameter SIL \cite{Baba99}. The excitation density enhanced by a factor of 2.83 at the center corresponds to a product of a convergence factor of $n^2$=3.40 and transmittance $[4n/(1+n)^2]^2$ = 0.83 for normal incidence at interfaces.

\begin{figure}
\includegraphics[width=.4\textwidth]{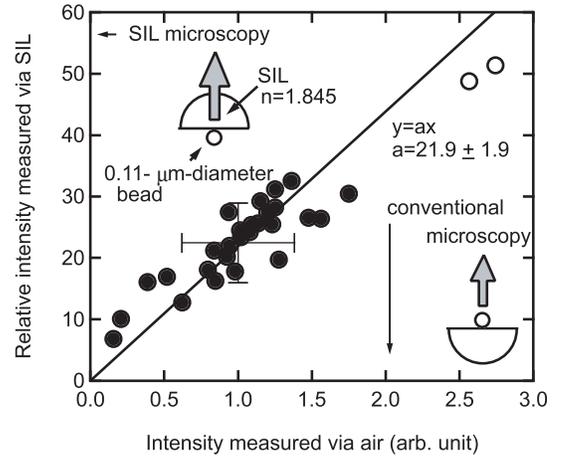}
\caption{Correlation plot of fluorescence intensities of 0.11-$\mu$m-diameter beads measured via air ($x$-axis) and via SIL with $n$=1.845 ($y$-axis). 
Solid and open circles indicate intensity of every individual bead. Straight line 
is a least-square fitting line for solid circles.}
\label{2}
\end{figure}

\begin{figure}
\includegraphics[width=.3\textwidth]{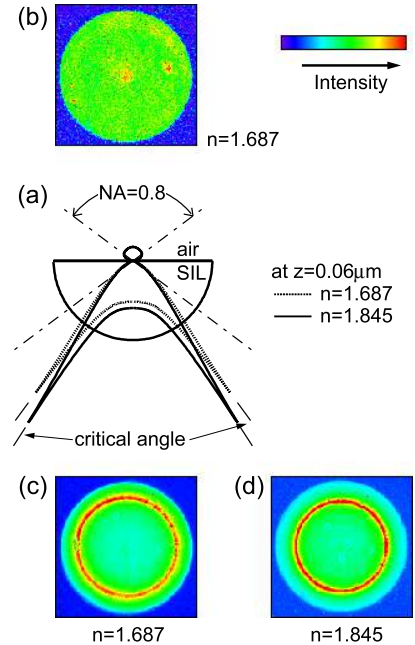}
\caption{(a) Calculated emission patterns from randomly oriented dipoles located in air at a distance $z$=0.06$\mu$m from the SIL-air surface. Dotted and solid lines are calculated for SILs of $n$=1.687 and 1.845, respectively.
(b) Fourier image of a 0.22-$\mu$m-diameter bead obtained for $n$=1.687 SIL 
and NA=0.8 objective via air. (c), (d) Fourier images obtained via SILs with $n$=1.687 and $n$=1.845, respectively.}
\label{3}
\end{figure}

\begin{figure}
\includegraphics[width=.3\textwidth]{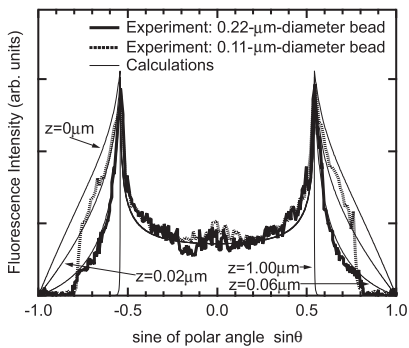}
\caption{Cross-sectional profiles of Fourier images across their centers in the SIL microscopy ($n$=1.845). Thick solid and dotted curves denote experimental profiles obtained with 0.22- and 0.11-$\mu$m-diameter beads, respectively. Four thin curves indicate calculated profiles for $z$=0, 0.02, 0.06, and 1.00 $\mu$m. All curves are normalized at $\sin\theta$=0.}
\label{4}
\end{figure}

Figure 2 shows fluorescence intensities of 29 beads with 0.11 $\mu$m diameter obtained with a NA=0.8 objective lens via SIL ($n$=1.845) and via air. Each data point corresponds to each bead. The $x$-axis represents the intensity measured in the conventional microscopy, while the $y$-axis indicates the intensity in the SIL microscopy. 
The distribution of data on the $x$-axis is wide, where $\bar{x}$ (mean value) $\pm$ $\sigma_x$ (standard deviation) of 1.0$\pm$0.38 is shown by the horizontal bar. Similarly wide distribution is observed on the $y$-axis, where $\bar{y} \pm \sigma_y$ of $22.4\times(1.0\pm0.29)$ is shown by the vertical bar.
The wide distribution is most likely caused by the size distribution of the beads, which is $\pm$30\% in volume, or  $\pm$10\% in diameter. 

The data are well fitted to a straight line $y=ax$, as shown in Fig. 2.  Two points far from the others, indicated by open circles, were omitted in the least-square fitting procedure, because each of them might be unresolved two or more beads. Then, the value of $a$, which represents the intensity ratio of the fluorescence observed in the SIL microscopy to the conventional one, is obtained as 21.9 with the estimated error as small as 1.9. 
To evaluate the emission intensity ratio of fluorescence for identical illumination density, we have to divide the ratio $a$ by the factor of $n^2[4n/(1+n)^2]^2$ = 2.83. The emission intensity ratio of 7.7$\pm$0.7 is obtained from $a=21.9\pm1.9$. 

Note that $\bar{y}/\bar{x}$ gives similar value of 22.4 for $a$ with a large uncertainty of 48\% as $[\sigma_x^2/\bar{x}^2 +\sigma_y^2/\bar{y}^2]^{1/2}$. While the correlation plot and the least-square fitting of Fig. 2 is useful to minimize the error of $a$, reasonably good estimation of $a$ is obtained simply by $\bar{y}/\bar{x}$. This supports the validity of our previous result for larger beads with 0.22 $\mu$m diameter, which was $a=13\pm$3.6 obtained as $\bar{y}/\bar{x}$ \cite{Koyama99}. The emission intensity ratio of 4.6$\pm$1.3 is obtained after division by 2.83. Upper columns in Table 1 summarize these results for two kinds of beads.

\begin{table}
\caption{
Summary of fluorescence intensity ratio and collection efficiency for a $n$=1.845 hemisphere SIL and a NA= 0.8 objective lens.
}
\begin{tabular}{l|ll}
\colrule
Bead diameter  ($\mu$m) & 0.11 &  0.22 \\
\colrule
Intensity ratio $a$ (Experiment) & 21.9$\pm$1.9 & 13$\pm$3.6\\
Efficiency ratio $a/n^2[4n/(1+n)^2]^2$ & 7.7$\pm$0.7 & 4.6$\pm$1.3 \\
\colrule 
Efficiency in the SIL microscopy & 62\% & 62\% \\
Efficiency in the conventional microscopy & 8\% & 14\% \\
\colrule
\end{tabular}
\end{table}

Figure 3 (a) shows examples of calculated emission patterns for a randomly oriented emission dipole located at a distance $z$ from the surface of a SIL with index $n$ assuming $z$=0.06 $\mu$m and $n$=1.845 (solid curve) and 1.687 (dashed curve) in the theoretical model of Ref. \cite{Hellen87} mentioned in the introduction. Note that the values of emission intensity ratio obtained by the above experiment correspond to the ratio of intensities integrated within $0<\sin\theta<$NA=0.8 in the SIL side and the air side, where $\theta$ is the polar angle of emission. 

To verify angle-dependent emission patterns like Fig. 3 (a), we next measured Fourier images of emission from the beads. In the measurement, we used the same Nikon optical microscope as the above measurement, but switched a tube lens to relay lenses and monitored fluorescence intensities on the back focal plane of the objective lens with a cooled CCD camera \cite{Feke98}. According to the Abbe's sine condition, the emission intensities in the direction of the polar angle $\theta$ from the beads are shown in the Fourier image intensities at the radius proportional to $\sin\theta$. 

Figure 3 (b) shows a Fourier image of a 0.22-$\mu$m-diameter bead on a SIL ($n$ = 1.687) obtained in the conventional microscopy configuration. The outline of the bright circular region represents the aperture of the objective, which corresponds to $\sin\theta_{\mbox{max}}$= 0.8 in this case. 
The Fourier image is fairly uniform within the circle, indicating that the emission depends little on the polar angle $\theta$ via air, similarly to the pattern in (a). 

Figure 3 (c) and (d) show Fourier images of a 0.22-$\mu$m-diameter bead obtained in the SIL microscopy with $n$=1.687 and 1.845, respectively. Each has a bright ring, which corresponds to the emission peak at the critical angle in the calculated pattern in (a). The ring diameter in (d) is obviously smaller than in (c). This is consistent with the difference of the calculated emission peak angle between $n$=1.687 and 1.845 as shown in (a). 

In order to examine the Fourier images, or the strong angular-dependence in emission patterns, in the SIL microscopy in detail, we plotted in Fig. 4 their cross-sectional profiles across their image centers, or $\sin\theta$ dependence of the emission intensity, for $n$=1.845 SIL. The thick solid and dotted curves in Fig. 4 show the cross-sectional profiles for the measured Fourier images of the emission from 0.22- and 0.11-$\mu$m-diameter beads, respectively, normalized at $\sin\theta$=0. Both curves go to zero for $\sin\theta > 0.8$ limited by NA of the objective lens. Remarkable difference between the two curves is observed in the region of $\sin\theta > 0.542=1/n$, or $\theta$ beyond the critical angle, since the light coupled via the evanescent field into the SIL is sensitive to the distance $z$ between the dipole and the SIL surface. The observed Fourier images demonstrate that the near-field optical coupling plays essential role in high collection efficiency in the SIL fluorescence microscopy.

For reference to the experimental data, we calculated emission patterns as functions of polar angle $\theta$ for randomly oriented emission dipoles located at various distance $z$ from the dielectric surface after Ref. \cite{Hellen87}. The four thin solid curves in Fig. 4 are cross-sectional profiles of Fourier images normalized at $\sin\theta$=0 calculated for $z$=0, 0.02, 0.06, and 1.00 $\mu$m. 
While all the four curves are similar for $\sin\theta < 0.542=1/n$, strong increase of intensity in $\sin\theta > 0.542=1/n$ occurs for smaller $z$. It is found that the two experimental curves for 0.22- and 0.11-$\mu$m-diameter beads are similar to the calculated curves for $z$=0.06 and 0.02 $\mu$m, respectively. 

The similarity between experimental data and calculation suggests that the fundamental process is modeled properly, though it is not understood why small values of $z$=0.06 and 0.02 $\mu$m gave the best fit for 0.22- and 0.11-$\mu$m-diameter beads, respectively, rather than the center-of-mass distances of 0.11 and 0.055 $\mu$m form the SIL surface. As a whole, however, the results of Fourier images in Figs. 3 and 4 together with the result of Fig. 2 confirm strong angular dependence in near-field optical coupling for efficient SIL fluorescence microscopy. 

In the end, we estimate absolute values of collection efficiency in the present experiments. For 0$<\sin \theta <$0.8, we indeed measured angular-dependent and angular-integrated intensities of fluorescence both in the air and SIL directions for 0.22- and 0.11-$\mu$m-diameter beads. If we extrapolate the angle-dependence data of two kinds of beads with the calculated curves for dipoles at $z$=0.02 and 0.06$\mu$m to the 0.8$<\sin \theta <$1 region, we complete emission patterns in all directions. 

Then the estimated values of collection efficiency in the SIL microscopy and in the conventional microscopy are 62\% and 14\% for 0.22-$\mu$m-diameter bead, whereas the values are 62\% and 8\% for 0.11-$\mu$m-diameter bead. The value of 62\% for the SIL microscopy is obtained by considering reflection loss at the spherical surface of the SIL. If we reduce the reflection loss by anti-reflection coating, the collection efficiency should be raised to 68\%. Remaining portions of 18\% and 24\% for 0.22- and 0.11-$\mu$m-diameter beads are lost in directions of 0.8$<\sin \theta <$1. These values, shortly summarized in lower columns of Table 1, show the significant advantage of the SIL in observing weak fluorescence of small samples efficiently. 

In summary, we observed strong angular dependence in fluorescence intensities in SIL fluorescence microscopy via Fourier imaging for 0.11- and 0.22-$\mu$m-diameter dye-doped polystyrene micro-sphere beads. 
The efficiency for 0.11$\mu$m diameter beads in SIL microscopy was 7.7$\pm$0.7 times higher than that measured in the conventional microscopy, and its absolute value was estimated to be 62\%. 
The results show that the efficient near-field optical coupling in SIL fluorescence microscopy are confirmed experimentally and are approximately modeled as dipole emissions near the SIL surface.

We would like to thank Dr. T. D. Harris (Praelux) for valuable discussion and suggestion. This work is partly supported by a Grant-in-Aid from the Ministry of Education, Science, Sports, and Culture, Japan.

\end{document}